\documentclass[preliminary,copyright]{eptcs}
\providecommand{\event}{FOCLASA 2012} 

\usepackage{amsmath}
\usepackage{amssymb}
\usepackage{amsthm}
\usepackage{paralist}
\usepackage{color}
\usepackage{mathrsfs}
\usepackage[width=0.7\linewidth, labelfont=bf, textfont={it}]{caption}
\usepackage{placeins}
\usepackage{multirow}
\usepackage{booktabs}
\usepackage{xspace}
\usepackage{amsthm}
\usepackage{listings}
\usepackage{algorithm}
\usepackage{algpseudocode}

\theoremstyle{definition}
\newtheorem{definition}{Definition}
\newtheorem{theorem}{Theorem}
\newtheorem{corollary}{Corollary}

\newcommand{\first}[1]{\textit{first}\left(#1\right)}
\newcommand{\last}[1]{\textit{last}\left(#1\right)}
\newcommand{\tr}[1]{\textit{tr}\left(#1\right)}

\newcommand{\sends}[1]{\overset{#1}{\longrightarrow}}
\newcommand{\paral}{\; \Vert \;}
\newcommand{\paralnospace}{\Vert}
\newcommand{\paraln}[1][1]{
	\overset{n}{\underset{i=#1}{\Vert}}}
\newcommand{\xor}{\; \oplus \;}
\newcommand{\xornospace}{\oplus}
\newcommand{\xorn}[1][1]{
		{\overset{n}{\underset{i=#1}{\bigoplus}}}}
\newcommand{\shuffle}{\; \otimes \;}
\newcommand{\shufflenospace}{\otimes}
\newcommand{\shufflen}[1][1]{
	\overset{n}{\underset{i=#1}{\bigotimes}}}
\newcommand{\seq}{\; \text{;} \;}
\newcommand{\seqnospace}{\text{;}}
\newcommand{\seqn}[1][1]{
	\overset{n}{\underset{i=#1}{\circledcirc}}}
\newcommand{\proj}{\; \triangleright \;}

\newcommand{\G}{\mathscr{G}}
\renewcommand{\L}{\mathscr{L}}

\newcommand{\D}{\Delta}

\newcommand{\systemA}{\text{System-A}\xspace}

\newcommand{\LEmpty}{\text{(L-Empty)}}
\newcommand{\LSeq}{\text{(L-Seq)}}
\newcommand{\LSeqN}{\text{(L-Seq-N)}}
\newcommand{\LSend}{\text{(L-Send)}}
\newcommand{\LRecv}{\text{(L-Recv)}}
\newcommand{\LChoice}{\text{(L-Choice)}}
\newcommand{\LChoiceN}{\text{(L-Choice-N)}}
\newcommand{\LParallel}{\text{(L-Parallel)}}
\newcommand{\LParallelN}{\text{(L-Parallel-N)}}
\newcommand{\LShuffle}{\text{(L-Shuffle)}}
\newcommand{\LShuffleN}{\text{(L-Shuffle-N)}}
\newcommand{\LExp}{\text{(L-Exp)}}
\newcommand{\LKleeneStar}{\text{(L-KleeneStar)}}
\newcommand{\LParen}{\text{(L-Paren)}}

\newcommand{\GEmpty}{\text{(G-Empty)}}
\newcommand{\GExp}{\text{(G-Exp)}\xspace}
\newcommand{\GSeq}{\text{(G-Seq)}}
\newcommand{\GSeqN}{\text{(G-Seq-N)}}
\newcommand{\GChoice}{\text{(G-Choice)}}
\newcommand{\GChoiceN}{\text{(G-Choice-N)}}
\newcommand{\GParallel}{\text{(G-Parallel)}}
\newcommand{\GParallelN}{\text{(G-Parallel-N)}}
\newcommand{\GShuffle}{\text{(G-Shuffle)}}
\newcommand{\GShuffleN}{\text{(G-Shuffle-N)}\xspace}
\newcommand{\GInteraction}{\text{(G-Interaction)}}
\newcommand{\GKleeneStar}{\text{(G-KleeneStar)}}
\newcommand{\GParen}{\text{(G-Paren)}}

\newcommand{\PInteraction}{\text{(P-Interaction)}\xspace}

\newcommand{\PExp}{\text{(P-Exp)}\xspace}
\newcommand{\PChoice}{\text{(P-Choice)}\xspace}
\newcommand{\PParallel}{\text{(P-Paral)}\xspace}
\newcommand{\PSeq}{\text{(P-Seq)}\xspace}
\newcommand{\PShuffle}{\text{(P-Shuffle)}\xspace}
\newcommand{\PSeqN}{\text{(P-Seq-N)}\xspace}
\newcommand{\PShuffleN}{\text{(P-Shuffle-N)}\xspace}
\newcommand{\PParalN}{\text{(P-Paral-N)}\xspace}
\newcommand{\PChoiceN}{\text{(P-Choice-N)}\xspace}

\def\defaultArrayStretch{1.2}

\renewcommand{\arraystretch}{\defaultArrayStretch}

\title{Parameterized Concurrent Multi-Party Session Types}
\author{Minas Charalambides
\email{charala1@illinois.edu}
\and
Peter Dinges
\email{pdinges@acm.org}
\and
Gul Agha
\email{agha@illinois.edu}
\and
\institute{Department of Computer Science\\
	University of Illinois at Urbana--Champaign, USA\\
	http://osl.cs.uiuc.edu/}
}
\def\titlerunning{Parameterized Concurrent Multi-Party Session Types}
\def\authorrunning{M. Charalambides, P. Dinges \& G. Agha}

\begin{document}
\maketitle

\begin{abstract}
  Session types have been proposed as a means of statically verifying
  implementations of communication protocols.  Although prior work has
  been successful in verifying some classes of protocols, it does not
  cope well with parameterized, multi-actor scenarios with inherent
  asynchrony.  For example, the sliding window protocol is
  inexpressible in previously proposed session type systems. This
  paper describes \systemA, a new typing language which overcomes many
  of the expressiveness limitations of prior work.  \systemA
  explicitly supports asynchrony and parallelism, as well as multiple
  forms of parameterization. We define \systemA and show how it can be
  used for the static verification of a large class of asynchronous
  communication protocols.
\end{abstract}

\section{Introduction}
\label{sec:intro}




Session types~\cite{DBLP:conf/esop/HondaVK98} are a means of
expressing the order of messages sent by
actors~\cite{DBLP:books/daglib/0066897} (or processes).  In
particular, session types can be used to statically check if a group
of processes communicate according to a given specification.  In these
systems, a \emph{global type} specifies the permissible sequences of
messages that participants may exchange in a given \emph{session}, as
well as the types of these messages.  The typing requires the
programmer to provide the \emph{global type}.  A \emph{projection}
algorithm then generates the restrictions implied by the global type
for each participant.  Such restrictions are called \emph{end-point
  types} or \emph{local types} and describe the expected behavior of
individual participants in the protocol.  The actual program code
implementing the participant behavior is checked for conformance
against this localized behavior specification.  We are interested in
generalizing prior work on session types to typing coordination
constraints on actors, which can then be enforced e.g.  with
\emph{Synchronizers}~\cite{DBLP:books/daglib/0015082,
  DBLP:conf/ecoop/FrolundA93, DBLP:conf/coordination/DingesA12} or
other ways~\cite{DBLP:journals/concurrency/MiliciaS05}.

This requires addressing two limitations of previous work.  First,
session types do not (directly) support \emph{asynchronous} events;
asynchronous communication leads to delays which require considering
arbitrary shuffles.  Second, we wish to consider \emph{parameterized}
protocols.  For example, consider two actors communicating through a
sliding window protocol: the actors agree on the length of the window
(i.e., the number of messages that may be buffered) and then proceed
to an exchange of concurrent messages.  Prior work on session types is
not suitable for typing protocols like this: the reason for this
deficiency is the fact that their respective type languages depend on
other formalisms for type checking (such as typed
$\lambda$-calculus~\cite{Barendregt92lambdacalculi} or System
T~\cite{godel1958systemT}) and these formalisms do not support a
parallel construct.

\paragraph{Contributions.}

We overcome many of these
shortcomings by developing \systemA, a new system for expressing types for multi-party
interaction that does not depend on other formalisms for type checking. 
The main contributions are
\begin{inparaenum}[(a)]
	\item 
		\emph{parameterized} constructs for
		expressing asynchrony, parallelism, sequence and
                choice (\autoref{sec:syntax}),
	\item 
		a projection mechanism to provide type constraints on
                individual actors (\autoref{sec:proj}),
	\item 
		the conditions under which conformance of the latter
		with the global type is assured 
								(Sections \ref{sec:realization}, \ref{sec:corr}) 
						and finally
	\item
		we show that structural equivalence of types
		is decidable in \systemA, by proving strong
                normalization of our local types (\autoref{sec:typechecking}).
\end{inparaenum}
Proofs of our theorems are included in the long version of this
paper~\cite{longVersion}.

\paragraph{Limitations.}

Using the strong normalization result, we can decide whether the local
behavior of an actor follows the protocol.  However, this result
relies on a type inference mechanism for the actor's behavior (of the
sort in Alur et al.~\cite{DBLP:conf/popl/AlurCMN05}).  We do not
describe such a type inference mechanism in this article.  Moreover,
we omit support for session delegation.  Finally, our realizability
results rely on structural criteria and are hence conservative rather
than precise~\cite{DBLP:conf/popl/BasuBO12}.





\section{Related Work}
\label{sec:related}






%

%

%

%
%

Session types~\cite{DBLP:conf/esop/HondaVK98,%
  DBLP:journals/entcs/YoshidaV07,DBLP:conf/parle/TakeuchiHK94,%
  DBLP:conf/concur/Honda93} originate from the context of
$\pi$-calculi as statically derivable descriptions of process
interaction behaviors.  In two-party sessions, they allow us to
statically verify that the participants have compatible behavior by
requiring \emph{dual} session types, that is, behaviors where each
participant expects precisely the message sequence that the other
participant sends and vice versa.  Extensions to session types support
asynchronous message passing~\cite{neubauer2004asynchronous} and
introduce subtyping~\cite{DBLP:journals/acta/GayH05} for a looser
notion of type compatibility.  Session types have been integrated into
functional~\cite{DBLP:journals/tcs/VasconcelosGR06,%
  DBLP:conf/haskell/PucellaT08} and object-oriented~\cite{%
  DBLP:conf/fmco/Dezani-CiancagliniGDY06,DBLP:conf/ecoop/HuYH08,%
  DBLP:conf/popl/GayVRGC10} languages among others,
with a wide range of applications including deadlock and livelock
detection~\cite{giachinodeadlocks}. Other extensions deal with
evolving system specifications using
transformations~\cite{eugster2012sound}. Exception handling, which
allows the participants of a protocol to escape the normal flow of
control and coordinate on another, has been considered in
\cite{DBLP:conf/sfm/CarboneYH09, DBLP:conf/concur/CarboneHY08}.  The
present article combines three enhancements to session types that
majorly extend their applicability: concurrent multi-party sessions,
parameterized session types, and an enhanced syntax.


\paragraph{Asynchronous Multi-Party Sessions.}

Many real-world protocols involve more than two participants, which
makes their description in terms of multiple two-party sessions
unnatural.  To overcome this limitation, Honda
et~al.~\cite{DBLP:conf/popl/HondaYC08} extend session types to support
multiple participants: a \emph{global type} specifies the interactions
between all participants from a global perspective.  A projection
algorithm then mechanically derives the behavior specification of the
individual participants, that is, the \emph{local type}.

%
%
%
%

The notion of global type and the associated correctness requirements
for projection were first studied by Carbone et
al.~\cite{DBLP:conf/esop/CarboneHY07}; Bonelli's work on multi-point
session types~\cite{DBLP:conf/tgc/BonelliC07} treats multi-party
protocols from the local perspective only.  Bettini et
al.~\cite{DBLP:conf/concur/BettiniCDLDY08} allow multi-party sessions
to interleave and derive a type system guaranteeing global progress.
Gay et al.~\cite{DBLP:journals/jfp/GayV10} consider subtyping 
in presence of asynchrony.

%
%
%

The present article builds on the foundation of a global protocol
specification and its projection onto local behaviors~\cite{DBLP:conf/popl/HondaYC08}.  However, we do not address the
question of local type safety and inference of actual programs, which
is a major part of Honda et al.'s work.  Furthermore, unlike their
approach (but following Castagna et
al.~\cite{DBLP:conf/forte/CastagnaDP11}), we simplify the notation for
global types by replacing recursion with the Kleene star and limiting
each pair of participants to use a single channel.
We introduce an explicit shuffle operator to preserve the
commutativity of message arrivals that can be achieved using multiple
channels.  Explicit shuffles also reduce the need for a special
subtyping relationship that allows the permutation of (Lamport-style)
concurrent asynchronous events for
optimization~\cite{DBLP:conf/esop/MostrousYH09}.

Following Castagna's global type syntax further, we support join
operations.  Joins cannot be expressed in Honda et~al.'s global types
because of the linearity requirement.  However, as Deni\'{e}lou
remarks~\cite{DBLP:conf/esop/DenielouY12}, join operations can only
describe series--parallel graphs.  Protocols such as the
\emph{alternating bit protocol} that require interleaved
synchronization between two processes consequently cannot be expressed
in our global type language.  Our choice to not support generic graph
structures as global types is founded on the desire to support
parameterization and, at the same time, keep the language
understandable; it remains unclear to us how to visualize
parameterized graphs in an intuitive fashion.

\paragraph{Parameterized Session Types.}

%
%
%

Our major extension of global types over Honda et al. and Castagna et
al.'s work is the introduction of parameters.  The starting point for
our parameterization of session types is the work of Yoshida
et~al.~\cite{DBLP:conf/fossacs/YoshidaDBH10} and
Bejleri~\cite{DBLP:conf/icfem/Bejleri10}.  Yoshida et~al. augment the
global types of Bettini et~al.~\cite{DBLP:conf/concur/BettiniCDLDY08}
with primitive recursive combinators to obtain dependent types that
support the parameterization of the repetition count and the
connection topology.  This allows, for example, using a single global
type for a highly participant-count dependent butterfly network.
Static verifiability---without instantiating the type parameters---is
maintained by projecting onto parameterized local types that allow
syntactic comparisons. In~\cite{DBLP:conf/popl/DenielouY11},
Deni{\'e}lou et~al. achieve parameterization by means of
quantification over behavior specifications they call \emph{roles}.
Like Bettini et al. and unlike \systemA, neither Deni{\'e}lou et~al. nor
Yoshida et~al.  support arbitrary, concurrency-induced shuffles
in their global and local types.  While Bettini et al. regain parallel
composition through the interleaving of global types, it is unclear
how the results transfer to the other two approaches.


\paragraph{Modeling of Multi-Party Protocols.}

%
%
%

%
%


Formalisms for describing multi-party communication protocols have
been studied in the context of designing distributed systems and
cryptographic protocols.  As modeling tools, the formalisms provide
ways to check a protocol for desired properties~\cite{yuang1988}, or
to synthesize such protocols~\cite{DBLP:journals/tc/ProbertS91}.  In
contrast to session types, the formalisms lack ways to statically
verify the compliance of an actual protocol implementation against the
specification.  Deni\'{e}lou and
Yoshida~\cite{DBLP:conf/esop/DenielouY12} discuss session types and
their relation to work on distributed systems or cryptographic
protocols in greater depth.

\section{Motivation}
\label{sec:motiv}

Formalisms introduced in previous work are not expressive enough to
define the types of some interesting protocols such as the sliding
window protocol, a locking--unlocking protocol, and a case of limited resource
sharing. In this section, we demonstrate how the behavior
of these protocols can be described in \systemA.

\paragraph{The Sliding Window Protocol.}

Assume an actor $a$ sends messages of type~$m$ to an actor $b$, which acknowledges
every received message with an $ack$ message. The protocol determines
that at most $n$ messages can be unacknowledged
at any given time, so that $a$ ceases sending until it receives another $ack$ message.
In this example, the
window size $n$ is a parameter, which means we need a way to express
the fact that $n$ sending--acknowledging events can be in transit at
any given instant in time.  Following is the global type of the
protocol.
\[
\underbrace{
\left( a \sends{m} b \seq b \sends{ack} a \right)^* \paral
\left( a \sends{m} b \seq b \sends{ack} a \right)^* \paral
\dots \paral
\left( a \sends{m} b \seq b \sends{ack} a \right)^*
}_{n \text{ times}}
\]
$a \sends{m} b$ denotes that $a$ sends a message of type $m$
to $b$. Operator $;$ is used for sequencing interactions. $\paralnospace$ is used for
composing its left and right arguments in parallel. The Kleene star has the usual
semantics and takes precedence over $\paralnospace$.

The above type can be expressed using the notation of Castagna et
al.~\cite{DBLP:conf/forte/CastagnaDP11}, albeit with a fixed window size~$n$.  
In \systemA on the other hand,
we can parameterize the type in~$n$ and
statically verify that participants follow the protocol without
knowing its value at runtime.
Using $\paraln$ to denote the parallel composition of $n$
processes, we obtain the following type in our notation:
\[
\paraln \left( a \sends{m} b \seq b \sends{ack} a \right)^*
\]

\paragraph{Locking / Unlocking.}

Consider a set of $n$ processes, each of which needs to acquire
exclusive access to a resource by sending it a $lock$ message. The
resource replies with $ack$, the process uses the resource and unlocks
it by sending an $unlock$ message, at which point the next process can
do the same, and so on. The following type describes the locking--unlocking
protocol for a fixed number of processes.
\[
	(c_1 \sends{lock} s \seq s \sends{ack} c_1 \seq c_1 \sends{unlock} s)
	\shuffle
	\dots
	\shuffle
	(c_{n} \sends{lock} s \seq s \sends{ack} c_{n} \seq c_{n} \sends{unlock} s)
\]
With $\shufflenospace$ denoting shuffling, this formula expresses that any
ordering of the
$(c_i \sends{lock} s \seq s \sends{ack} c_i \seq c_i \sends{unlock} s)$
events is acceptable.
To support a dynamic network topology, the number of participants should be
a parameter. The following is the locking--unlocking example in \systemA, where
conformance to
the protocol is statically verifiable without knowledge of the runtime value of~$n$.
\[
	\shufflen	(c_i \sends{lock} s \seq s \sends{ack} c_i \seq c_i \sends{unlock} s)
\]

\paragraph{Limited Resource Sharing.}

In this scenario, a server $s$ grants two clients $c_1$ and $c_2$ 
exclusive access to a set of $n$ resources.
At any given point, a maximum of $n$ resources can be locked, but the relevant
lock--ack--unlock messages from both clients can be interleaved in any way.
Following is the global type for this situation:
\[
	\paraln
		\left(
		c_1 \sends{lock_i} s \seq s \sends{ack_i} c_1 \seq c_1 \sends{unlock_i} s
		\xor
		c_2 \sends{lock_i} s \seq s \sends{ack_i} c_2 \seq c_2 \sends{unlock_i} s
		\right)^*
\]
The parallel composition is parameterized in~$n$, the number of resources.
Each sequence of lock--ack--unlock messages is also parameterized in $i$, which ranges
from 1 to $n$. This is necessary to ensure realizability of the protocol, as in the case of multiple
outstanding requests, it allows the participants to disambiguate the responses
they receive. Each parallel instance subsumed by the $\paraln$ operator consists of a loop (Kleene Star) which
entails a choice, indicated by $\xornospace$. Either $c_1$ gets access to a resource,
or $c_2$ and this happens repeatedly.

\section{Type Syntax}
\label{sec:syntax}


\subsection{\label{sub:globaltypes}Global Types}

A global type describes
a protocol which the whole system must adhere to. The examples in \autoref{sec:motiv}
are all global types, since they describe the behavior of all participants. Global
types in \systemA can be constructed according to the grammar in \autoref{tab:global-types},
with operator descriptions following.

\begin{table}[h!tb]
	\centering
	\caption{The grammar of global types}
	\label{tab:global-types}
	\begin{tabular}{lllllll}
		\toprule
		$\G::=$ &&$(\G)$ &\GParen &\textbar &$\epsilon$ &\GEmpty\\
		 &\textbar &$\G \seq \G$ &\GSeq &\textbar &$\seqn\G_i$ &\GSeqN\\
		 &\textbar &$\G \xor \G$ &\GChoice &\textbar &$\xorn \G_i$ &\GChoiceN\\
		 &\textbar &$\G \paral \G$ &\GParallel &\textbar &$\paraln\G_i$ &\GParallelN\\
		 &\textbar &$\G \shuffle \G$ &\GShuffle &\textbar &$\shufflen\G_i$ &\GShuffleN\\
		 &\textbar &$p_1 \sends{type} p_2$ &\GInteraction &\textbar &$\G^n$ &\GExp\\
		 &&&&\textbar &$\G^*$ &\GKleeneStar\\
	 \bottomrule
	\end{tabular}
\end{table}

\begin{description}
	\item[\GSeq] is used for the sequential composition of events.
	\item[\GChoice] denotes exclusive choice between the arguments. 
		For instance, $\G_1 \xor \G_2$ means that
		either $\G_1$ or $\G_2$ will be executed (but not both).
	\item[\GParallel] means that the arguments run in parallel; any interleaving
		of sequenced actions is possible. For instance, 
		$(a \sends{t_1} b \seq a \sends{t_2} c) \paral c \sends{t_3} b$
		means that any of the interleavings ABC, ACB, CAB is possible, where
		A = $(a \sends{t_1} b)$, B = $(a \sends{t_2} c)$
		and C = $(c \sends{t_3} b)$. Notice that B is not allowed to precede A, as the
		ordering of actions as determined by operator $\seqnospace$ is not allowed to change.
	\item[\GShuffle] means that both arguments are executed atomically, in an unspecified order.
		Formally,	$\G_1 \shuffle \G_2 \equiv (\G_1 \seq \G_2) \xor (\G_2 \seq \G_1)$.
	\item[\GInteraction] denotes the sending and receiving of a message.
		For instance,
		$p_1 \sends{t} p_2$ means that process $p_1$ sends a message
		of type $t$ to process $p_2$.
	\item[\GKleeneStar] has the usual semantics, of zero or more repetitions of the argument.
\end{description}

The $n$-ary versions of the operators express
behaviors where the value of $n$ is unknown at compile time.  
\GSeqN, \GChoiceN, \GParallelN, \GShuffleN apply the respective binary
operator $n-1$ times to $n$ global types, parameterized in~$i$.
\GExp denotes $n$-fold repetition of the argument (in sequence).
Note that for known values of~$n$, we do not need the right
column of \autoref{tab:global-types}, as the desired
behavior can be produced by suitable repeated applications of
the binary operators.

All of the operators are commutative, with the exception of
sequencing.  All operators are furthermore associative, with the
exception of shuffling.  In particular,
\[
\shufflen \G_i \neq (\dots(\G_1 \shuffle \G_2) \shuffle \G_3 \dots ) 
\shuffle \dots \shuffle \G_n).
\]
Instead, $\shufflen \G_i$ means that all arguments $\G_i$ are
executed atomically, but in an unspecified order.

The distinction between the Kleene star and exponentiation is fundamental.
The use of $\G^n$ means that the protocol conformance checker will have 
to prove that the system
is correct for any fixed value of the parameter $n$. $\G^*$ on the other hand means
an arbitrary number of repetitions of $\G$. There is no parameter fixing this number
and it may be different from instance to instance of the Kleene Star and/or
among executions of the same program with the same run-time values for its parameters.

\subsection{\label{sub:localtypes}Local Types}

A local type specifies the abstract behavior of a single protocol
participant.  The syntax of local types is given in
\autoref{tab:local-types}, with descriptions following.

\begin{table}[h!tb]
	\centering
	\caption{The grammar of local types}
	\label{tab:local-types}
	\begin{tabular}{rllllll}
	\toprule
	$\L::=$ &&$(\L)$ &\LParen &\textbar &$\epsilon$ &\LEmpty\\
	 &\textbar &$a ! t$ &\LSend &\textbar &$a ? t$ &\LRecv\\
	 &\textbar &$\L \seq \L$ &\LSeq &\textbar &$\seqn \L_i$ &\LSeqN\\
	 &\textbar &$\L \xor \L$ &\LChoice &\textbar &$\xorn (\L_i)$ &\LChoiceN\\
	 &\textbar &$\L \paral \L$ &\LParallel &\textbar &$\paraln \L_i$ &\LParallelN\\
	 &\textbar &$\L \shuffle \L$ &\LShuffle &\textbar &$\shufflen \L_i$ &\LShuffleN\\
	 &\textbar &$\L^n$ &\LExp &\textbar &$\L^*$ &\LKleeneStar\\
	\bottomrule
	\end{tabular}
\end{table}

\begin{description}
	\item[\LSeq, \LChoice, \LParallel, \LShuffle, \LExp,
          \LKleeneStar] are defined as in the case of global types
		(\autoref{sub:globaltypes}).

                With \LParallel{} being defined as in the global case, the local type
                $(a ! t \seq a ! u) \paral a ? v$ again allows three orderings of the
                events $T=a!t$, $U=a!u$, and $V=a?v$: $TUV$, $TVU$, and $VTU$.  As
                above, the specification $a!t \seq a!u$ enforces that $T$ happens
                before~$U$.

	\item[\LSend] denotes sending a message of type $t$ to process $a$.
	\item[\LRecv] denotes receiving a message of type $t$ from process $a$.
\end{description}

In the sliding window example of \autoref{sec:motiv}, the behavior of the sender $a$ is
described by the local type $\paraln (b ! m \seq b ? \text{ack})^*$.
Leaving out the initial $\paraln$ symbol for the time being, what remains is
$(b ! m \seq b ? \text{ack})^*$. This means 
sending a message and then receiving an acknowledgment ($b ! m \seq b ? \text{ack}$), an
arbitrary number of times.
Assuming that the window size $n$ is a parameter, any interleaving of $n$ of these
sequences is possible, with the obvious constraint of not receiving more
acknowledgments than the number of messages sent. This is ensured by composing sequences
of the form ($b !m \seq b ? \text{ack}$), where ordering is forced by the $\seqnospace$ operator.



\section{Projection}
\label{sec:proj}

The local type of the sliding window protocol in
\autoref{sub:localtypes} is a restriction of the respective global
type in \autoref{sec:motiv} onto the individual processes.  In this
section, we investigate a way of automating this process.  $\G \proj
p$ is read ``the projection of global type $\G$ onto process $p$'' and
the result is a local type as defined in \autoref{tab:local-types}.
The projection function $\proj$ is formally defined in
\autoref{tab:proj-def} and the result of applying it to all the
processes in the system is an \emph{environment} $\Delta = \{p_i \, :
\, \L_i\}_{i \in I}$ which maps processes to local types.

\renewcommand{\arraystretch}{1.8}

\begin{table}[h!tb]
	\centering
	\caption{The projection function}
	\label{tab:proj-def}
	\begin{tabular}{rlll}
		\toprule
		$(a \sends{m} b) \proj p$ &::= &$\begin{cases}b ! m &\text{if } p=a\\
																						a ? m &\text{if } p=b\\
																						\epsilon &\text{otherwise}\end{cases}$ &\PInteraction\\
		 $\G^n \proj p$ &::= & $(\G \proj p)^n$ &\PExp\\
		 $(\G_1 \xor \G_2) \proj p$ &::= & $(\G_1 \proj p) \xor (\G_2 \proj p)$ &\PChoice\\
		 $(\G_1 \paral \G_2) \proj p$ &::= & $(\G_1 \proj p) \paral (\G_2 \proj p)$ &\PParallel\\
		 $(\G_1 \seq \G_2) \proj p$ &::= & $(\G_1 \proj p) \seq (\G_2 \proj p)$ &\PSeq\\
		 $(\G_1 \shuffle \G_2) \proj p$ &::= & $(\G_1 \proj p) \shuffle (\G_2 \proj p)$ &\PShuffle\\
		 $(\seqn \G_i) \proj p$ &::= & $\seqn (\G_i \proj p)$ &\PSeqN\\
		 $(\xorn \G_i) \proj p$ &::= & $\xorn (\G_i \proj p)$ &\PChoiceN\\
		 $(\shufflen \G_i) \proj p$ &::= & $\shufflen (\G_i \proj p)$ &\PShuffleN\\
		 $(\paraln \G_i) \proj p$ &::= & $\paraln (\G_i \proj p)$ &\PParalN\\
		\bottomrule
	\end{tabular}
\end{table}

\renewcommand{\arraystretch}{\defaultArrayStretch}

For the lock/unlock example of \autoref{sec:motiv},
projecting onto a client $c_k$ and the server $s$ yields
\begin{align*}
\G \proj c_k &=	\shufflen (c_i \sends{lock} s \seq s \sends{ack} c_i \seq c_i \sends{unlock} s \proj c_k) &\PShuffleN\\
						 &= \underset{i \neq k}{\bigotimes}\epsilon \shuffle (s ! lock \seq s ? ack \seq s ! unlock) &\PInteraction, \PSeq\\
						 &= s ! lock \seq s ? ack \seq s ! unlock, &\text{(eliminating $\epsilon$)}
\\
\G \proj s &=	\shufflen (c_i \sends{lock} s \seq s \sends{ack} c_i \seq c_i \sends{unlock} s \proj s) &\PShuffleN\\
					 &= \shufflen (c_i ? lock \seq c_i ! ack \seq c_i ? unlock)\,. &\PInteraction, \PSeq
\end{align*}
Similarly, the projected local types for the resource sharing protocol 
of \autoref{sec:motiv} are
\begin{align*}
	\L_s &= \paraln
		\left(
		c_1 ? lock_i \seq c_1 ! ack_i \seq c_1 ? unlock_i
		\xor
		c_2 ? lock_i \seq c_2 ! ack_i \seq c_2 ? unlock_i
		\right)^*\\
	\L_{c_1} &= \paraln
		\left(
		s ! lock_i \seq s ? ack_i \seq s ! unlock_i
		\right)^*\\
	\L_{c_2} &= \paraln
		\left(
		s ! lock_i \seq s ? ack_i \seq s ! unlock_i
		\right)^* \,.
\end{align*}

\section{Type Checking}
\label{sec:typechecking}

Given a global type, we need to be able to check the respective projections
against the local types inferred from the program itself.
This is possible due to the following properties of our language of
local types:

\theoremstyle{plain}

\begin{theorem}[Weak Normalization]
	For any local type $\L$ in \systemA, there exists a finite
	sequence of reduction steps which brings the type to a normal form.
\end{theorem}

We prove this in the extended version of this paper~\cite{longVersion}, 
where we provide the reduction semantics
and a normalization process.

\begin{corollary}[Strong Normalization]
	For every local type $\L$ in \systemA, all sequences
	of reduction steps are finite and lead to the same normal form.
\end{corollary}

In the extended version of this paper, we show that the aforementioned normalization process 
uniquely determines the reduction semantics, implying the uniqueness of normal forms.

Checking structural equivalence of the types derived from the program against the projections 
is decidable up to $\alpha$-conversion. However,
all that is required to overcome this issue is that names in the code are consistent with those
in the supplied global type.  In our opinion it is reasonable to
expect programmers to adhere to such a naming convention.

\section{Global Type Realization}
\label{sec:realization}

In this section, we discuss the properties that a given global type must satisfy in order
to be \emph{projectable}. These properties are discussed while assuming 
actor semantics~\cite{DBLP:books/daglib/0066897}
for the messaging system; that is, 
asynchronous, unordered and eventual (guaranteed, albeit with arbitrary delay) 
delivery of messages. 
Applying the projection function to a \emph{projectable}
global type will result in local types for the participants, whose
combined behavior is consistent with the global type---a fact we show
in \autoref{sec:corr}.

The subsequent discussion of projectability criteria uses the
following definitions:

\theoremstyle{definition}
\begin{definition}[Event]
        An \emph{event} is a single interaction $p_1 \sends{m} p_2$ in a global type.

        We extend the projection function onto events and
	write $e \proj p$ to denote the projection of event $e$ onto process $p$
	using rule \PInteraction.
\end{definition}

\begin{definition}[Trace]
	A \emph{trace} is a sequence of events producible by a global type $\G$ and
	is of the form $e_1 \seq e_2 \seq \dots \seq e_k$.
	The set of traces a global type $\G$ can produce is denoted by $\tr{\G}$.
	The first and last events of a trace $t$ are denoted $\first{t}$ and $\last{t}$
	respectively. Abusing notation, the set of events that appear
        first in traces of~$\G$
				is denoted $\first{\G} = \{\first{t} \mid t\in
				\tr{\G}\}$. Similarly, the set of events that appear last in
				traces of~$\G$ is denoted $\last{\G}$.
	
	Since a trace is simply a sequence of events of the form $p \sends{m} q$, we
	extend the projection function onto traces in the natural way.
	We write $t \proj p$ to denote the projection of
	trace $t$ onto process $p$ using rules \PSeq and \PInteraction.
\end{definition}

\subsection{Sequentiality Criterion}
\label{sub:seqCriterion}

The purpose of this criterion is to ensure that the sequential
constructs of a global type retain sequential semantics after
projection. As an example problematic case, consider $\G_1 = a
\sends{m_1} b \seq c \sends{m_2} d$.  Without the use of some covert
coordination channel (for example by implementing a barrier
mechanism), it is impossible for $c$ to know when $b$ has received the
message. The two events $a \sends{m_1} b$ and $c \sends{m_2} d$ are
impossible to order using our projection function, as the resulting
environment would be 
$\D_1 = \{a : b ! m_1, \, b : a ? m_1, \, c : d ! m_2, \, d : c ? m_2\},$ 
which allows $c$ to send
$m_2$ to~$d$ before $a$ sends $m_1$ to~$b$. $\G_1$ does not
satisfy the sequentiality criterion and thus is not projectable. 

Another problematic case is $\G_2 = a \sends{m_1} b \seq a \sends{m_2} b$, where
$a$ cannot know when $m_1$ has been received so as to start transmitting $m_2$,
hence $\G_2$ is not projectable either.
The following definition captures the conditions under which events
are guaranteed to respect the sequencing restrictions imposed in a global type,
when the latter is projected onto individual processes.

\begin{definition}[Sequentially Projectable Global Type]
	The set of \emph{sequentially projectable} ($SP$) global types is
	defined inductively as follows:
	\[
	\begin{cases}
		p_1 \sends{t} p_2 \quad\in SP \quad\forall p_1, p_2 \in \Pi\\
		p_1 \sends{m_1} p_2 \seq p_2 \sends{m_2} p_3 \quad\in SP \quad\forall p_1, p_2, p_3 \in \Pi\\
		\\
		\big(\forall e_1 \in \last{\G_1}, e_2 \in \first{\G_2} \Rightarrow (e_1 \seq e_2) \in SP \big)
			&\Rightarrow \quad\G_1 \seq \G_2 \in SP\\
		\big(\forall e_1 \in \last{\G_i\{1/i\}}, e_2 \in \first{\G_i\{2/i\}} \Rightarrow (e_1 \seq e_2) \in SP \big)
			&\Rightarrow \quad \seqn\G_i \in SP\\
		\big(\forall e_1 \in \last{\G}, e_2 \in \first{\G} \Rightarrow (e_1 \seq e_2) \in SP \big)
			&\Rightarrow \quad\G^n \in SP\\
		\big(\forall e_1 \in \last{\G}, e_2 \in \first{\G} \Rightarrow (e_1 \seq e_2) \in SP \big)
			&\Rightarrow \quad\G^* \in SP
	\end{cases}
	\]
	where $\Pi$ denotes the set of processes. 
\end{definition}

Illustrating the third case of the definition above, the following global type is in $SP$:
\[
	\G = (a \sends{m} b \paral c \sends{m} b) \seq (b \sends{m} l \paral b \sends{m} k)
\]
It is easy to see that 
$\last{a \sends{m} b \paral c \sends{m} b} = \{a \sends{m} b, \; c \sends{m} b\}$
and
$\first{b \sends{m} l \paral b \sends{m} k}$\\
$= \{b \sends{m} l, \; b \sends{m} k\}$
so that all four sequences (e.g. $a \sends{m} b \seq b \sends{m} k$) 
are in $SP$ according to the first two lines of the definition above.

\subsection{Choice Criterion}
\label{sub:choiceCriterion}

The purpose of this criterion is to ensure that projecting
$\G_1 \xor \G_2$ maintains the choice semantics, meaning that
all participants can recognize which branch of the choice operator
they need to take during execution. As an example of a type that does
not satisfy this criterion, consider
\[
	\G = (a \sends{m_1} b \seq b \sends{k} c \seq c \sends{t_1} d)
		\xor (a \sends{m_2} b \seq b \sends{k} c \seq c \sends{t_2} d).
\]
Here, $a$ and $b$ know which branch they are on, because on the left
branch $b$ receives a message of type $m_1$ from $a$, while on the
branch on the right it receives a message of type $m_2$. However,
from that point on, $b$ behaves identically with respect to $c$,
which has no way of telling whether the message to send to
$d$ should be of type $t_1$ or $t_2$. We call the first point at which
two traces differ with respect to a given process the \emph{distinctive point},
which can be $\epsilon$ if no such point exists. This notion is formalized
in the following definition:

\begin{definition}[Distinctive Point]
	The \emph{distinctive point} of a process $p$ with respect
	to a pair of traces $t_1 = \left(e_1, \dots, e_{k}\right) \in \tr{\G}$ 
	and $t_2 = \left(f_1, \dots, f_{l}\right) \in \tr{\G}$ is
	an index $i$ given by
	\[
		d_{t_1, t_2}(p) =
			\min \{ i \mid (e_i \proj p) \neq (f_i \proj p) \}
	\]
	where $e_j, f_j$ denote events.
	In the case where $t_1 \proj p = \epsilon$, or $t_2 \proj p = \epsilon$, or
	$t_1 \proj p = t_2 \proj p$, no such $i$ exists and the distinctive point is
	defined to be $\epsilon$.
\end{definition}

The definition that follows captures the conditions under which the choice
semantics are maintained after projection. The first bullet deals with the non
parameterized version of the choice operator~$\xornospace$. Item (i) captures the case where
a process $p$ is the first process acting on the two branches, in which case 
it must inform the others
of the branch they are on. It does so by either sending a different message,
or by sending to a different process in each case. Note that the same process
must inform the others on both branches.
Item (ii) captures the case where $p$ is not the first
process to act, in which case it must be informed of the branch it
is on and the distinctive point should be a suitable
receive event.

Notice how the second bullet deals with shuffling by means of choice.
Clearly, if a process can tell whether it is on $\G_1$ or $\G_2$,
it is also able to tell the order in which they appear.

The third bullet inductively uses the previous two to define choice-wise projectability
in the parameterized cases of choice~$\xornospace$ and shuffle~$\shufflenospace$. 
\begin{definition}[Choice-Wise Projectable Global Type]
	The set of 
	\emph{Choice-Wise Projectable} ($CP$) global types is defined
        inductively as follows:
	\begin{itemize}
		\item
			$\G = \G_1 \xor \G_2 \in CP$ \emph{iff} 
			$\forall p \in \G$, either of the following is true:
			\begin{align*}
				\text{(i)}\quad
					&\G = \G_1 \xor \G_2 \text{ and}\\
					&\forall e_1 \in \first{\G_1}, \; e_2 \in \first{\G_2} \; : \;
						e_1 = p \sends{m} q, \; e_2 = p \sends{m'} q'\quad
						&\text{where } p \neq q \text{ and } p \neq q'\\
						&&\text{and }	(q \neq q' \text{ or } m \neq m')\\
				\text{(ii)}\quad
					&\forall t_1 = \left(s_1,\dots,s_{k_1}\right)\in \tr{\G_1}, 
					\; t_2 = \left(u_1,\dots,u_{k_2}\right) \in \tr{\G_2},\\
						&\text{either } d_{t_1, t_2}(p) = \epsilon\text{, or }\\
					&d_{t_1, t_2}(p) = i\text{ and }
						s_i = q \sends{m} p, \; u_i = q' \sends{m'} p
						&\text{where } p \neq q \text{ and } p \neq q'\\
						&&\text{and }	(q \neq q' \text{ or } m \neq m')
			\end{align*}
		\item
			$\G = \G_1 \shuffle \G_2 \in CP\quad$ \emph{iff}
			$\quad\G_1 \xor \G_2 \in CP$
		\item
				$\G = \xorn \G_i \in CP\quad$ \emph{iff} $\quad(\G_i\{1/i\} \xor \G_i\{2/i\}) \in CP$\\
				$\G = \shufflen \G_i \in CP\quad$ \emph{iff} $\quad(\G_i\{1/i\} \shuffle \G_i\{2/i\}) \in CP$
	\end{itemize}
\end{definition}
The criterion for parameterized shuffling~$\shufflen$ is stricter than what one can derive if the value of $n$ is given. 
However, it is hard to loosen up the constraint when it is dealt with as a parameter.

\subsection{Parallel Composability Criterion}
\label{sub:paralCriterion}

As an example of what can go wrong when composing two global types
using the $\paralnospace$ operator, consider the example
$\G = \left((a \sends{m_1} b \seq b \sends{k_1} c) \xor (a \sends{m_2} b \seq b \sends{k_2} c)\right)
\paral a \sends{m_1} b$.
The intended behavior of $\G$ is that $a$ chooses whether to send a message of
type $m_1$ or $m_2$ to $b$, which in turn decides whether to send $c$ a message
of type $k_1$ or $k_2$. Concurrently with this, an additional $m_1$ is sent
from $a$ to $b$. Assume that as far as $\xornospace$ is concerned, $a$ decides to send $m_2$
to $b$. It is then obvious how 
the additional parallel event $a \sends {m_1} b$ might confuse $b$ to simultaneously take
both branches of the choice operator.

In general, the problem appears when actions in one parallel branch affect choices made on another.
Global types that do not exhibit this problem are \emph{parallel projectable} $(PP)$.

\begin{definition}[Parallel Projectable Global Types]
	For two global types $\G_1$ and $\G_2$, $\G = \G_1 \paral \G_2$
	is parallel projectable ($PP$) if there is no overlap between the distinctive points
	in $\G_1$ and events in $\G_2$.
	Formally, 
	\begin{itemize}
		\item
			$\G_1 \paral \G_2 \in PP\quad$ \emph{iff} 
			$\quad\forall t_1 = (e_1,\dots,e_k), \; t_1' = (e'_1,\dots,e'_{k'}) \in \tr{\G_1}, \; t_2 \in \tr{\G_2}, \; p \in \Pi \;$ we have 
				$d_{t_1, t_1'}(p) = i$ and one of the following is true:
				\begin{align*}
					\text{(a)}\quad
						&i = \epsilon\\
					\text{(b)}\quad
						&e_i, e'_i \text{ both have $p$ as the sender}\\
					\text{(c)}\quad
						&e_i \notin t_2 \text{ and }	e'_i \notin t_2
				\end{align*}
		\item
			$\paraln \G_i \in PP\quad$ \emph{iff} $\quad (\G_i\{1/i\} \paral \G_i\{2/i\}) \in PP$
	\end{itemize}
	where $\Pi$ denotes the set of processes.
	Notice how this definition incorporates
	parallel composability of two Kleene starred types (the Kleene Star entails a
	choice pertaining to loop entrance and exit).
\end{definition}

\subsection{Kleene Star Criterion}
\label{sub:kleeneCriterion}

Use of the Kleene Star in global types can result in protocols
whose projection is unsafe, that is, can result in execution
traces that are not part of the original global type. To avoid
this, a global type must be such that the entry and exit conditions
to the starred type can be identified by all participants.
Determining whether this is the case requires inspection
of not only the starred type itself, but also of what comes after
the starred section.

\begin{definition}[Kleene Star Projectable Global Types]
	For global types $\G, \G'$, we say that
	$\G^* \seq \G'$ is Kleene Star Projectable $(KP)$
	\emph{iff} $\G \xor \G' \in CP$.
\end{definition}

As an example of a type that is not in $KP$, consider
$\G = (a \sends{m} b \seq b \sends{m'} c)^* \seq c \sends{m''} d$
where $c$ has no way of knowing whether it should wait for $m'$ from $b$,
or proceed immediately with sending $d$ the message $m''$.

\section{Correctness}
\label{sec:corr}

The conditions discussed above are
sufficient to ensure that the projection function generates local types
which are functionally consistent with the global type.  We call a global type
that satisfies all of the above criteria
\emph{projectable}:

\begin{definition}[Projectable Global Type]
	The set of \emph{projectable} ($PR$) global types is inductively defined in \autoref{tab:projectable}.

	\begin{table}[h!tb]
		\caption{The set $PR$}
		\label{tab:projectable}
		\centering
		\begin{tabular}{lllll}
			\toprule
			$\epsilon \in PR$\\
			$a \sends{m} b \in PR$\\
			$\G \in PR\quad$ and $\quad\G^n \in SP$ &$\Rightarrow$ &$\G^n \in PR$\\
			$\G_1, \G_2 \in PR\quad$ and $\quad(\G_1 \seq \G_2) \in SP$ &$\Rightarrow$ &$(\G_1 \seq \G_2) \in PR$\\
			$\G_1, \G_2 \in PR\quad$ and $\quad(\G_1 \xor \G_2) \in CP$ &$\Rightarrow$ &$(\G_1 \xor \G_2) \in PR$\\
			$\G_1, \G_2 \in PR\quad$ and $\quad(\G_1 \shuffle \G_2) \in CP$ &$\Rightarrow$ &$(\G_1 \shuffle \G_2) \in PR$\\
			$\G_1, \G_2 \in PR\quad$ and $\quad(\G_1 \paral \G_2) \in PP$ &$\Rightarrow$ &$(\G_1 \paral \G_2) \in PR$\\
			$\G_1, \G_2 \in PR\quad$ and $\quad\G_1^* \in SP\quad$ and $\quad(\G_1^* \seq \G_2) \in SP \cap KP$ &$\Rightarrow$ &$(\G_1^* \seq \G_2) \in PR$\\
			$\G_i\{1/i\}, \G_i\{2/i\} \in PR\quad$ and $\quad(\G_i\{1/i\} \seq \G_i\{2/i\}) \in SP$ &$\Rightarrow$ &$\seqn \G_i \in PR$\\
			$\G_i\{1/i\}, \G_i\{2/i\} \in PR\quad$ and $\quad(\G_i\{1/i\} \xor \G_i\{2/i\}) \in CP$ &$\Rightarrow$ &$\xorn \G_i \in PR$\\
			$\G_i\{1/i\}, \G_i\{2/i\} \in PR\quad$ and $\quad(\G_i\{1/i\} \xor \G_i\{2/i\}) \in CP$ &$\Rightarrow$ &$\shufflen \G_i \in PR$\\
			$\G_i\{1/i\}, \G_i\{2/i\} \in PR\quad$ and $\quad(\G_i\{1/i\} \paral \G_i\{2/i\}) \in PP$ &$\Rightarrow$ &$\paraln \G_i \in PR$\\
			\bottomrule
		\end{tabular}
	\end{table}
\end{definition}

\autoref{th:projectable} formalizes our intuition that under the constraints mentioned above, the projection function is correct; that is,
the projected environment is consistent with the global type.
In what follows, $\tr{\Delta}$ with $\D = \{p_i : \L_i\}_{i \in I}$ denotes the set of traces producible
by environment $\D$. Also, $\Delta_{\G}$ denotes the environment resulting from the projection of $\G$ onto
the set of processes, i.e. $\Delta_{\G} = \{p : \G \proj p\}_{p \in \Pi}$.

\theoremstyle{plain}
\begin{theorem}
	\label{th:projectable}
	$\G \in PR \Rightarrow \tr{\G} = \tr{\Delta_\G}$
\end{theorem}

We sketch the proof of this theorem in the extended version of this paper~\cite{longVersion}, 
where we inductively treat each of the cases in \autoref{tab:projectable}. 
What needs to be proved is essentially
$\forall t \in \tr{\G} \Leftrightarrow t \in \tr{\Delta_{\G}}$. While
the forward direction is rather obvious, proving that the
projected environment does not generate traces that are not part of the original global type
is trickier and is why we need the criteria of \autoref{sec:realization}. 

Proving this theorem,
we get a correctness proof of our projection function (given the premises discussed
previously) for free.

\section{Conclusions and Future Work}
\label{sec:conclusion}

We introduced \systemA which allows for parameterized parallelism,
where the number of participants, the types of messages sent, as well
as the number of such messages are controlled by type parameters.
Choice among various execution paths can also be parameterized, so
that the number and types of different paths to be taken is not known
at compile time.  \systemA also introduces a shuffling operator, which
expresses arbitrary reordering of its arguments, again in a
parameterized fashion.  A series of examples demonstrates the
usefulness of these extensions, which allow us to specify and check
previously inexpressible interactions such as the sliding window
protocol and parallel resource locking/unlocking
(\autoref{sec:motiv}).  In \systemA, we can statically
verify---without instantiating the parameters---the compliance of
implementations to protocols: we do this by first projecting
(\autoref{sec:proj}) the specification to parameterized types, and
then comparing these projections against the types extracted from the
program.  An important result we obtain is that structural equivalence
of types in \systemA is decidable; we present this result in
\autoref{sec:typechecking} by first showing weak and subsequently 
strong normalization of local types.  Unlike other typing
proposals, \systemA does not depend on other theories (typed
$\lambda$-calculus, system~T, or system~F) for type-checking.  In
\autoref{sec:realization} we discuss the conditions under which our
projection function is correct and state their sufficiency
in \autoref{sec:corr}.

\paragraph{Future Work.}

Complete type checking with \systemA is only decidable up to type
inference; we do not provide an algorithm for inference of suitable
types in an actor language.  The design of a programming language
along with the relevant type inference algorithm is the next step
towards a practical implementation.

Another practical consideration includes semantic comparison of local
types.  Our normalization algorithm~\cite{longVersion} already
includes many cases of semantically equivalent, yet structurally
differing types. Semantic comparison is unnecessary for the weak
normalization proof, but would be useful in a practical setting where
the user is interested in semantic adherence to a protocol.
Specifically for the case where reordering of terms is possible as a
result of operator commutativity, our suggested coding only serves as
an existential proof.  A more practical coding scheme could be
developed, perhaps employing lexicographic ordering.

Deni{\'e}lou et al.~\cite{DBLP:conf/popl/DenielouY11} propose a system
where parameterization is achieved by means of quantification over
\emph{roles}.  Roles are behavior specifications that are taken up by
processes while they participate in a protocol, and processes are
allowed to join and leave protocols (respectively, adopt and drop
roles) dynamically.  Their notation's expressiveness is
limited when it comes to arbitrary, concurrency-induced interleavings
of events. Nevertheless, incorporating their ideas in \systemA would greatly expand
the applicability of the ideas presented here, towards
a different direction than what is addressed in the present paper.

Support for session delegation and exception handling (in the sense of
Carbone et al.~\cite{DBLP:conf/concur/CarboneHY08}) represents another
opportunity for extension.  Furthermore, it may be possible to
transfer the recent, precise realizability
results~\cite{DBLP:conf/popl/BasuBO12} for
choreographies~\cite{DBLP:journals/computer/Peltz03} to our
parameterized specifications.

\section*{Acknowledgments}

The authors would like to thank the anonymous reviewers for their
insightful comments.  This publication was made possible in part by
sponsorships from the Air Force Research Laboratory and the Air Force
Office of Scientific Research under agreement number FA8750-11-2-0084,
as well as the Army Research Office under Award No. W911NF-09-1-0273.
The U.S. Government is authorized to reproduce and distribute reprints
for Governmental purposes notwithstanding any copyright notation thereon.

\FloatBarrier
\bibliographystyle{eptcs}
\bibliography{session-types}

\end{document}